\documentclass[12pt,a4paper]{article}
\pdfoutput=1

\usepackage[utf8]{inputenc}
\usepackage[T1]{fontenc}
\usepackage[margin=2.5cm]{geometry}
\usepackage{amsmath,amssymb}
\usepackage{graphicx}
\usepackage{booktabs}
\usepackage{array}
\usepackage{xcolor}
\usepackage{hyperref}
\usepackage{natbib}
\usepackage{setspace}
\usepackage{caption}
\usepackage{fancyhdr}
\usepackage{abstract}
\usepackage{titlesec}
\usepackage{textgreek}

\newcommand{\Msun}{M_{\odot}}
\newcommand{\Mearth}{M_{\oplus}}

\hypersetup{
    colorlinks=true,
    linkcolor=blue!60!black,
    citecolor=blue!60!black,
    urlcolor=blue!60!black,
    pdftitle={Mass inventory of the Solar System beyond the Sun},
    pdfauthor={Mario Menichella},
}

\titleformat{\section}{\large\bfseries}{\thesection.}{0.5em}{}
\titleformat{\subsection}{\normalsize\bfseries}{\thesubsection}{0.5em}{}
\titleformat{\subsubsection}{\normalsize\itshape}{\thesubsubsection}{0.5em}{}

\newcolumntype{L}[1]{>{\raggedright\arraybackslash}p{#1}}
\newcolumntype{C}[1]{>{\centering\arraybackslash}p{#1}}
\newcolumntype{R}[1]{>{\raggedleft\arraybackslash}p{#1}}

\begin{document}

\begin{center}
	
{\fontsize{19}{22}\bfseries
	Mass Inventory of the Solar System Beyond the Sun:\\[4pt]
	A Systematic Compilation with Uncertainty Budget
}
	
	\vspace{16pt}
	
	{\normalsize
		Mario Menichella$^{1}$
	}
	
	\vspace{4pt}
	
{\small
	$^{1}$\textit{William Herschel Astronomical Observatory, Apuan Alps, Stazzema (LU), 55040, Italy}
}
	
	\vspace{12pt}
	
	{\small
		E-mail: m.menichella@gmail.com
	}
	
	\vspace{10pt}
	
{\normalsize March 2026}
	
\end{center}

\vspace{8pt}
\noindent\rule{\textwidth}{0.8pt}

\begin{abstract}
\noindent
We compile a systematic mass inventory of the Solar System
excluding the Sun, drawing on spacecraft measurements,
planetary ephemerides, and population surveys of small-body
populations including main-belt asteroids, trans-Neptunian
objects, and long-period comets as tracers of the Oort cloud,
to assemble a single reference table with propagated
uncertainties for all major components. Using a Monte
Carlo simulation with $10^5$ realisations, and treating poorly
constrained components (scattered disc, Oort cloud) as log-normal
distributions, we obtain a total non-solar mass of $462\,\Mearth$
(median), with a 68\% credible interval of $[451,\,515]\,\Mearth$ and a
90\% credible interval of $[449,\,642]\,\Mearth$. The giant planets
dominate with $444.6\,\Mearth$ (96.2\% of the total). A variance
decomposition shows that 98.2\% of the total uncertainty in the
current Solar System mass budget is attributable
to a single component: the inner Oort cloud (Hills cloud), for which no
direct observational constraints exist. The current small-body
populations retain only $\sim$0.2\% of the primordial trans-Neptunian
disc mass inferred from Nice Model simulations, and $\sim$0.04\% of the
primordial asteroid belt mass implied by the Grand Tack hypothesis. We
identify constraints on the Oort cloud from the Vera C.\ Rubin
Observatory and improved long-period comet surveys as the primary path
toward a better-determined total mass budget.

\medskip
\noindent\textbf{Keywords:} Solar System --- mass inventory --- Oort
cloud --- Kuiper belt --- planetary formation
\end{abstract}

\noindent\rule{\textwidth}{0.8pt}
\vspace{6pt}

\section{Introduction}

The Solar System is, by mass, almost entirely the Sun. The Sun contains
approximately 99.86\% of the total mass of the system, with everything
else --- eight planets, hundreds of moons, millions of asteroids,
billions of trans-Neptunian objects, and a vast reservoir of cometary
nuclei in the outer reaches --- contributing the remaining 0.14\%. Yet
it is precisely this remaining fraction that records the history of the
Solar System's formation and dynamical evolution, and that determines
most of its observable complexity.

Surprisingly, no single published work provides a comprehensive,
updated, and self-consistent inventory of this non-solar mass, complete
with uncertainty budgets for each component. Individual component masses
are scattered across dozens of papers in planetary science, celestial
mechanics, and Solar System dynamics: planetary masses in ephemeris
papers \citep{Park2021}, the asteroid belt mass from dynamical analyses
of planetary residuals \citep{Pitjeva2018}, the Kuiper belt from
dedicated survey papers \citep{Petit2023}, and the Oort cloud from
long-period comet statistics and dynamical models \citep{Brasser2008,
Francis2005}. These results use different units, different reference
epochs, and different definitions of component boundaries, making direct
comparison and summation non-trivial. Review articles on Solar System
structure discuss mass budgets qualitatively, but do not assemble a
quantitative inventory with propagated uncertainties.

This paper attempts to fill that gap. Our goal is simple: to compile,
from the current literature, the best available mass estimate for each
major component of the Solar System excluding the Sun, to express all
results in a common unit (Earth masses, $\Mearth$), and to propagate
the uncertainties honestly and systematically. The result is a single
reference table that can serve as a baseline for future studies.

Three considerations motivate this work beyond mere bookkeeping.

The first is the extreme dynamic range of component masses and their
uncertainties. The planets are known to better than one part in a
million; the Oort cloud is uncertain by two orders of magnitude. A
systematic inventory makes this contrast visible and quantitative in a
way that no individual paper has done before, and identifies clearly
where observational effort would most improve our knowledge of the Solar
System's total mass budget.

The second is the relevance of the current mass inventory to models of
Solar System formation. Both the Nice Model \citep{Tsiganis2005,
Gomes2005} and the Grand Tack hypothesis \citep{Walsh2011} predict
specific amounts of mass depletion from the primordial asteroid belt and
trans-Neptunian disc as a result of giant planet migration and dynamical
instability. An accurate current inventory, compared with these
theoretical predictions, provides a quantitative test of these formation
scenarios and a baseline for assessing how much mass was lost and to
where.

The third motivation is the growing relevance of Solar System mass
comparisons in the context of exoplanetary science. ALMA observations of
protoplanetary discs around other stars have made it possible to
estimate the mass available for planet formation in other systems, and
comparative planetology increasingly requires knowing how our own
system's mass distribution compares with the range observed elsewhere.

The structure of this paper is as follows. Section~\ref{sec:method}
describes our methodology. Sections~\ref{sec:planets}
through~\ref{sec:oort} treat each component in turn. Section~\ref{sec:results}
presents the integrated mass inventory and Monte Carlo uncertainty
analysis. Section~\ref{sec:discussion} discusses the results in the
context of Solar System formation models and exoplanetary disc
comparisons. Section~\ref{sec:conclusions} summarises our conclusions.

Throughout this paper we use Earth masses ($\Mearth = 5.972 \times
10^{24}$\,kg) as the primary unit. Solar masses ($\Msun = 1.989 \times
10^{30}$\,kg) are used where appropriate. All masses refer to the
present epoch.

\section{Methodology}
\label{sec:method}

Our approach is a literature compilation with systematic uncertainty
propagation. We do not derive new observational constraints, nor do we
construct new dynamical models. Instead, we identify, for each
component of the Solar System, the most precise and up-to-date mass
estimate available in the refereed literature, assign it a central value
and an uncertainty, and combine the components into a total mass budget
with a consistently propagated uncertainty.

\subsection{Source selection}

For each component we prioritise, in order: (i) direct
spacecraft-derived measurements, where available; (ii) dynamical
estimates from high-precision planetary ephemerides; (iii) population
models calibrated against observational surveys; (iv) purely theoretical
estimates from formation and dynamical evolution models. The first two
categories apply to the planets, major moons, and the main asteroid
belt, and yield uncertainties below 1\%. The third category applies to
the Kuiper belt and scattered disc, with uncertainties of order
20--50\%. The fourth category applies exclusively to the Oort cloud,
where uncertainties span one to two orders of magnitude.

Where multiple independent estimates exist for the same component, we
adopt the most recent measurement consistent with the majority of
published values, and note significant discrepancies explicitly.

\subsection{Uncertainty characterisation}

We treat uncertainties differently depending on their nature and
magnitude. For well-determined components (planets, major moons,
asteroid belt), uncertainties are approximately Gaussian and small. We
quote the $1\sigma$ uncertainty as reported in the source paper and
propagate them in quadrature:
\begin{equation}
    \sigma^2_\mathrm{tot} = \sum_i \sigma^2_i.
\end{equation}
For components with large and asymmetric uncertainties --- specifically
the scattered disc, the inner Oort cloud, and the outer Oort cloud --- a
Gaussian description is inappropriate. We use a log-normal distribution,
characterised by a multiplicative factor $f$ such that the 68\%
credible interval spans $[M/f,\; M \cdot f]$, where $M$ is the central
estimate.

\subsection{Monte Carlo uncertainty propagation}

To obtain the uncertainty on the total mass we use a Monte Carlo
simulation with $N = 10^5$ realisations. We draw each component mass
from its assigned distribution (Gaussian for well-constrained
components, log-normal for poorly constrained ones), sum the components
in each realisation, and report the resulting distribution. The
simulation is implemented in Python using NumPy and can be readily reproduced from the distributions and parameters described above. The median, 16th percentile, and 84th percentile
of the resulting distribution are our reported central value and
$1\sigma$ equivalent bounds.

\subsection{Unit conventions and scope}

All masses are expressed in Earth masses ($\Mearth = 5.972 \times
10^{24}$\,kg). We include all gravitationally bound objects for which
published mass estimates exist at the time of writing (early 2025). We
explicitly exclude interplanetary dust and gas ($<10^{18}$\,kg,
negligible at our level of precision) and the hypothetical Planet Nine,
discussed separately in Section~\ref{sec:p9}. No attempt is made to
account for the time evolution of the mass budget.

\section{Component Mass Estimates}

\subsection{The Planets}
\label{sec:planets}

The masses of the eight planets are among the most precisely determined
quantities in astronomy, derived from spacecraft tracking data and
gravitational perturbations encoded in high-precision planetary
ephemerides. The data used here are from the JPL planetary ephemeris
DE440 \citep{Park2021}, currently the most comprehensive dynamical model
of the Solar System.

Planetary masses in units of $10^{24}$\,kg are as follows: Mercury
0.330, Venus 4.87, Earth 5.97, Mars 0.642, Jupiter 1898, Saturn 568,
Uranus 86.8, Neptune 102. In units of $\Mearth$, the gas and ice giants
dominate overwhelmingly: Jupiter alone accounts for $317.8\,\Mearth$,
followed by Saturn ($95.2\,\Mearth$), Neptune ($17.1\,\Mearth$), and
Uranus ($14.5\,\Mearth$). The four terrestrial planets combined
contribute only $\sim$$2.07\,\Mearth$.

The total planetary mass is therefore $\sim$$447.7\,\Mearth$, of which
the giant planets account for $444.6\,\Mearth$ --- approximately 99.3\%
of all planetary mass. Jupiter alone accounts for 71\% of the total
planetary mass and is 2.5 times more massive than all other planets
combined. The uncertainties on all planetary masses are below 0.01\% and
are entirely negligible compared to the uncertainties on the
trans-Neptunian populations.

\subsection{Natural Satellites and Ring Systems}

The Solar System hosts more than 290 confirmed natural satellites. For
this inventory we treat them in three tiers: the large planetary-mass
moons whose masses are individually well-determined, the mid-sized moons
with masses known to $\sim$1--10\%, and the small irregular satellites
whose combined mass is negligible.

\subsubsection{Large planetary-mass moons}

The seven largest moons each exceed the mass of Pluto. Their individual
masses are given in Table~\ref{tab:moons}.

\begin{table}[h]
\centering
\caption{Masses of the seven largest moons of the Solar System
(spacecraft-derived, uncertainty $<0.1$\%).}
\label{tab:moons}
\begin{tabular}{lccc}
\toprule
\textbf{Moon} & \textbf{Parent planet} & \textbf{Mass ($10^{22}$\,kg)} & \textbf{Mass ($\Mearth$)} \\
\midrule
Ganymede  & Jupiter & 14.82 & 0.02482 \\
Titan     & Saturn  & 13.45 & 0.02253 \\
Callisto  & Jupiter & 10.76 & 0.01803 \\
Io        & Jupiter &  8.932 & 0.01496 \\
Moon      & Earth   &  7.346 & 0.01230 \\
Europa    & Jupiter &  4.800 & 0.00804 \\
Triton    & Neptune &  2.147 & 0.00360 \\
\midrule
Subtotal  & ---     & 62.25  & 0.104   \\
\bottomrule
\end{tabular}
\end{table}

\subsubsection{Mid-sized moons and rings}

The $\sim$20 additional satellites with diameters exceeding 200\,km
(Titania, Oberon, Rhea, Iapetus, and others) contribute a combined
$\sim$$0.004\,\Mearth$. Saturn's rings have a total mass of $(1.54 \pm
0.49) \times 10^{19}$\,kg \citep{Iess2019}, equivalent to $\sim$$2.6
\times 10^{-6}\,\Mearth$. All other ring systems are at least two
orders of magnitude less massive. The combined satellite mass is
$\sim$$0.108\,\Mearth$, dominated by the seven large moons.

\subsection{The Main Asteroid Belt}

The main asteroid belt occupies the region between approximately 2.06
and 3.27\,AU. Its total mass is known with comparatively good precision
from planetary ephemeris analyses, which measure the total gravitational
perturbation exerted by the belt on the orbits of Mars and the inner
planets. The main asteroid belt also acts as the long-term source region
for most near-Earth asteroids (NEAs) through collisional fragmentation
and resonance-driven transport processes (see e.g., \citealt{Menichella1996}).

\subsubsection{The dominant objects}

The four largest objects --- Ceres, Vesta, Pallas, and Hygiea ---
contain an estimated 62\% of the belt's total mass, with Ceres alone
accounting for 39\%. The Dawn mission determined the mass of Ceres as
$(939.3 \pm 0.5) \times 10^{18}$\,kg and the mass of Vesta as
$(259.076 \pm 0.001) \times 10^{18}$\,kg. Together the top four objects
sum to approximately $1.49 \times 10^{21}$\,kg.

\subsubsection{Total belt mass}

\citet{Pitjeva2014, Pitjeva2018}, using the EPM ephemerides and
approximately 800,000 positional observations, modelled the asteroid
belt as a two-dimensional homogeneous annulus spanning 2.06 to 3.27\,AU.
The consensus total belt mass is approximately $2.39 \times 10^{21}$\,kg,
or $\sim$$4 \times 10^{-4}\,\Mearth$ ($\pm$20\%). Jupiter Trojans,
Hildas, and NEAs contribute a further $\sim$$4.4 \times
10^{-5}\,\Mearth$ combined.

The primordial main belt is estimated to have been 150--250 times more
massive than today, implying an original mass of $\sim$0.06--0.10\,$\Mearth$.
More than 99.6\% of the original material has been lost through
dynamical ejection driven by Jupiter's gravitational perturbations.

\subsection{The Kuiper Belt and Trans-Neptunian Region}

The trans-Neptunian region (30--1000\,AU) hosts several dynamically
distinct populations. Unlike the Oort cloud, most have been partially
surveyed directly.

\subsubsection{The classical Kuiper belt}

The most precise mass estimate from a direct observational survey is
provided by \citet{Petit2023} using the full OSSOS dataset, who find a
total mass of $0.014\,\Mearth$ for non-resonant main-belt objects
brighter than $H_r = 8.3$. An independent dynamical approach using the
EPM2017 ephemerides \citep{Pitjeva2018} finds $(1.97 \pm 0.30) \times
10^{-2}\,\Mearth$. We adopt $0.020\,\Mearth$ ($\pm$30\%) combining both
approaches.

\subsubsection{Scattered disc and detached objects}

The scattered disc (highly eccentric orbits with perihelia near
30--35\,AU) contains between 240,000 and 830,000 objects larger than
$\sim$18\,km in diameter. Total mass is estimated at $\sim$$0.05\,\Mearth$
(factor $\sim$3 uncertainty). The detached/sednoid population (Sedna,
2012~VP$_{113}$) is estimated at $<$$0.01\,\Mearth$.

\subsubsection{Planet Nine (hypothetical)}
\label{sec:p9}

If Planet Nine exists with a mass of $\sim$5--10\,$\Mearth$ at
$\sim$400--800\,AU \citep{Batygin2016}, it would represent a significant
addition to the inventory. Its existence is unconfirmed; we treat it as
a separately labelled speculative component and exclude it from our
total.

The total trans-Neptunian mass excluding the Oort cloud is of order $0.08-0.1\,\Mearth$, comparable to roughly one-tenth of an Earth mass and about 6–8 times the mass of the Moon.

\subsection{The Oort Cloud}
\label{sec:oort}

The Oort cloud represents by far the largest source of uncertainty in
any mass inventory of the Solar System. Its existence is inferred from
the orbital distribution of long-period comets \citep{Oort1950}, and
every mass estimate is therefore model-dependent.

\subsubsection{Structure and observational basis}

The Oort cloud is divided into two regions. The outer Oort cloud
($\sim$20,000--100,000\,AU) is approximately spherical and subject to
perturbations from passing stars and galactic tides. The inner Oort
cloud, or Hills cloud \citep{Hills1981}, is a denser disc-shaped
structure extending from roughly 2,000 to 20,000\,AU. The sole
observational constraint on the cloud's mass comes from the observed
flux of long-period comets. Estimating the total mass requires three
uncertain assumptions: (1) the total population of cometary nuclei; (2)
the size-frequency distribution of nuclei; (3) the mass of a typical
nucleus. Each is uncertain by at least a factor of a few, and they
multiply.

\subsubsection{Evolution of mass estimates}

Table~\ref{tab:oort} summarises published mass estimates for the outer
Oort cloud. Early high estimates (\citealt{Marochnik1988}: $\sim$$100\,\Mearth$;
\citealt{Weissman1983}: $\sim$$38\,\Mearth$) assumed Halley's comet is
representative. Improved knowledge of the long-period comet size
distribution led to drastic downward revision. Modern N-body formation
models converge on $\sim$$1\,\Mearth$ for the outer cloud alone;
observationally-based estimates suggest $\sim$3--5\,$\Mearth$.

\begin{table}[h]
\centering
\caption{Published mass estimates for the outer Oort cloud.}
\label{tab:oort}
\begin{tabular}{lll}
\toprule
\textbf{Reference} & \textbf{Method} & $\boldsymbol{M_\mathrm{Oort}}$ \textbf{($\Mearth$)} \\
\midrule
\citet{Weissman1983}        & LPC flux + Halley proxy          & $\sim$38          \\
\citet{Marochnik1988}       & LPC flux + angular momentum      & $\sim$100         \\
\citet{Weissman1996}        & Revised size distribution        & $\sim$1.9         \\
\citet{SternWeissman2001}   & N-body + collisional evolution   & $\sim$1           \\
\citet{Francis2005}         & Updated LPC catalogue            & $\sim$3--4        \\
\citet{Brasser2008}         & N-body simulation                & $0.75 \pm 0.25$   \\
\citet{Fernandez2000}       & Formation model                  & $1.0 \pm 0.4$     \\
\bottomrule
\end{tabular}
\end{table}

\subsubsection{The inner Oort cloud (Hills cloud)}

The Hills cloud presents an additional compounding uncertainty. No
published mass estimate based on direct observational constraints exists
for the inner Oort cloud. Dynamical models predict it to be
significantly more massive than the outer cloud, with model-based
estimates of 0.1--10\,$\Mearth$ from formation simulations. Taking the
outer cloud as $\sim$1--5\,$\Mearth$ and the Hills cloud as 1--5$\times$
that value, the total Oort cloud spans an estimated range of roughly 2
to 100\,$\Mearth$, with a central value near $\sim$$10\,\Mearth$.

\subsubsection{Adopted values}

We adopt a three-tier representation: lower bound $1\,\Mearth$ (outer
cloud only, \citealt{Brasser2008}); central estimate $10\,\Mearth$
(outer $\sim$3--5\,$\Mearth$ + Hills cloud $\sim$5--10\,$\Mearth$); upper
bound $100\,\Mearth$ \citep{Marochnik1988}. This three-order-of-magnitude
range exceeds the combined mass of all other small-body populations.

In the Monte Carlo implementation (see Table~3), this broad interval is represented by modeling the outer and inner Oort components as independent log-normal distributions. The adopted medians ($3.00\,\Mearth$ and $6.89\,\Mearth$, respectively) and multiplicative $1\sigma$ factors are selected to reproduce the dynamical plausibility range discussed above, while consistently propagating asymmetric uncertainties.

\section{Results: Mass Inventory and Uncertainty Analysis}
\label{sec:results}

\subsection{Component-by-component summary}

Table~\ref{tab:inventory} presents the complete mass inventory derived
from the component estimates of Section~3 and the Monte Carlo procedure
of Section~\ref{sec:method}. All values are the median of $N = 10^5$
Monte Carlo realisations, with the 16th and 84th percentiles as
$1\sigma$ equivalent bounds.

The relative masses of the minor
Solar System components and their uncertainties are shown on a logarithmic scale in
Figure 1.

\begin{table}[h]
\centering
\caption{Complete mass inventory of the Solar System excluding the Sun.
The 90\% credible interval for the total is $[448.5,\,642.0]\,\Mearth$.}
\label{tab:inventory}
\setlength{\tabcolsep}{5pt}
\begin{tabular}{L{5.2cm} C{1.8cm} C{1.8cm} C{1.8cm} L{2.8cm}}
\toprule
\textbf{Component} & \textbf{Median ($\Mearth$)} & \textbf{16th \%} & \textbf{84th \%} & \textbf{Notes} \\
\midrule
Giant planets              & 444.60     & 444.59     & 444.61     & JPL DE440 \\
Terrestrial planets        & 2.070      & 2.069      & 2.071      & JPL DE440 \\
Major moons (7 largest)    & 0.1040     & 0.1039     & 0.1041     & Spacecraft tracking \\
Mid-sized moons            & 0.0040     & 0.0038     & 0.0042     & Jacobson et al. \\
Small satellites + rings   & $5\times10^{-6}$ & $3\times10^{-6}$ & $7\times10^{-6}$ & Iess et al.\ 2019 \\
Main asteroid belt         & $4.0\times10^{-4}$ & $3.2\times10^{-4}$ & $4.8\times10^{-4}$ & Pitjeva \& Pitjev 2018 \\
Trojans + Hildas + NEAs    & $4.4\times10^{-5}$ & $2.2\times10^{-5}$ & $6.6\times10^{-5}$ & Estimate \\
Classical Kuiper belt      & 0.0200     & 0.0140     & 0.0260     & Petit et al.\ 2023 \\
Dwarf planets (excl.\ Ceres, Eris) & 0.0027 & 0.0024 & 0.0030 & Direct masses \\
Scattered disc             & 0.050      & 0.017      & 0.147      & Log-normal, $f=3$ \\
Detached / sednoids        & 0.005      & 0.001      & 0.025      & Log-normal, $f=5$ \\
Oort cloud (outer)         & 3.00       & 0.60       & 14.9       & Log-normal, $f=5$ \\
Oort cloud (inner/Hills)   & 6.89       & 0.99       & 47.8       & Log-normal, $f=7$ \\
\midrule
\textbf{TOTAL}             & \textbf{462.4} & \textbf{450.8} & \textbf{515.4} & --- \\
\bottomrule
\end{tabular}
\end{table}

\subsection{Structural features of the inventory}

The four giant planets account for 96.2\% of the median total. Jupiter
alone contributes 68.7\% of the entire non-solar mass. All small-body
populations combined, excluding the Oort cloud, amount to approximately
$0.08\,\Mearth$ --- less than the mass of Mars.

The Oort cloud median contribution (outer + inner) is $\sim$$9.9\,\Mearth$,
making it the third-largest component after the giant planets and the
terrestrial planets. The variance decomposition shows that 98.2\% of the
total variance in the Monte Carlo distribution comes from the inner Oort
cloud alone. This is the central quantitative result of this paper:
\textit{the total mass of the Solar System excluding the Sun is uncertain
almost entirely because of the Oort cloud.}

The variance contribution of each component was estimated by fixing all other variables at their median values and computing the resulting conditional variance of the total mass. 
Within this framework, the inner Oort cloud accounts for $\approx 98\%$ of the total variance budget.

\subsection{Monte Carlo distribution}

\begin{figure}[h]
\centering
\includegraphics[width=0.85\textwidth]{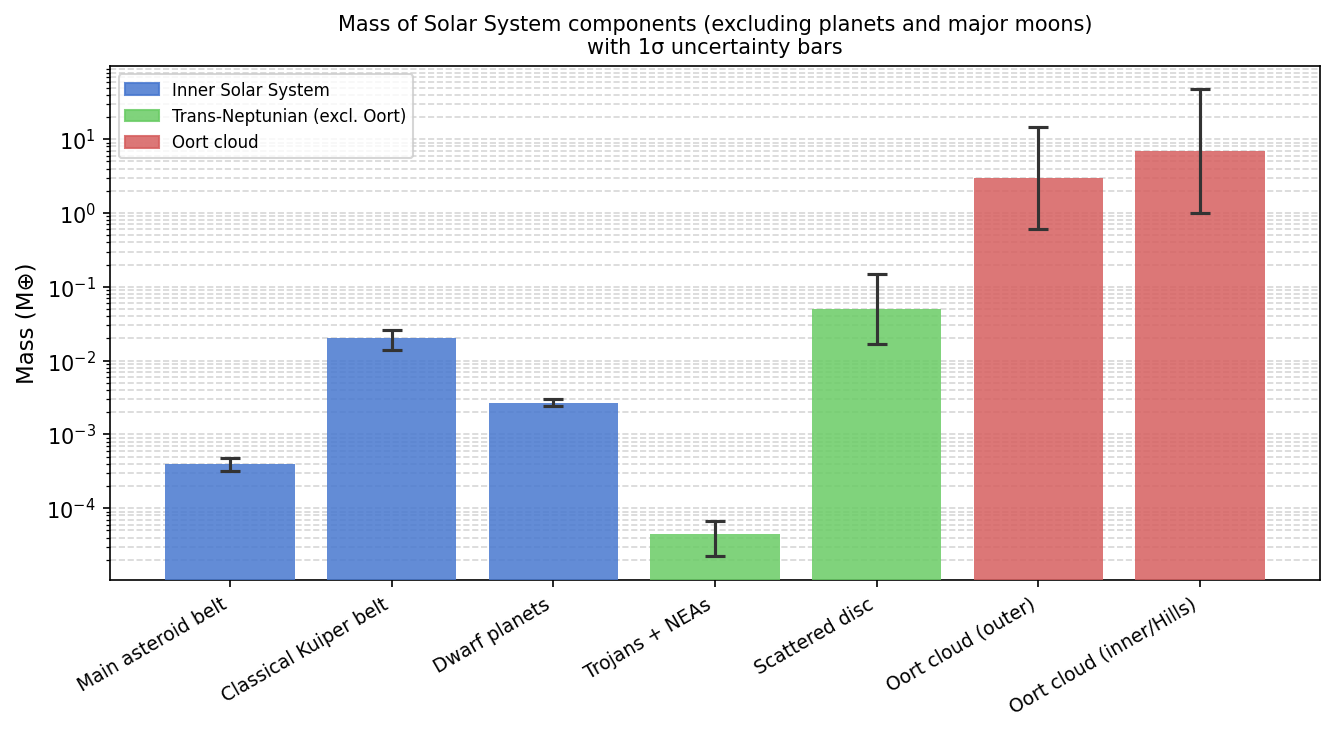}
\caption{Mass of Solar System components (excluding planets and major
moons) on a logarithmic scale, with $1\sigma$ uncertainty bars. Blue:
inner Solar System populations. Green: trans-Neptunian populations
(excluding Oort cloud). Red: Oort cloud components. The Oort cloud
error bars span more than one order of magnitude, dominating the total
uncertainty budget.}
\label{fig:components}
\end{figure}

\begin{figure}[h]
\centering
\includegraphics[width=0.85\textwidth]{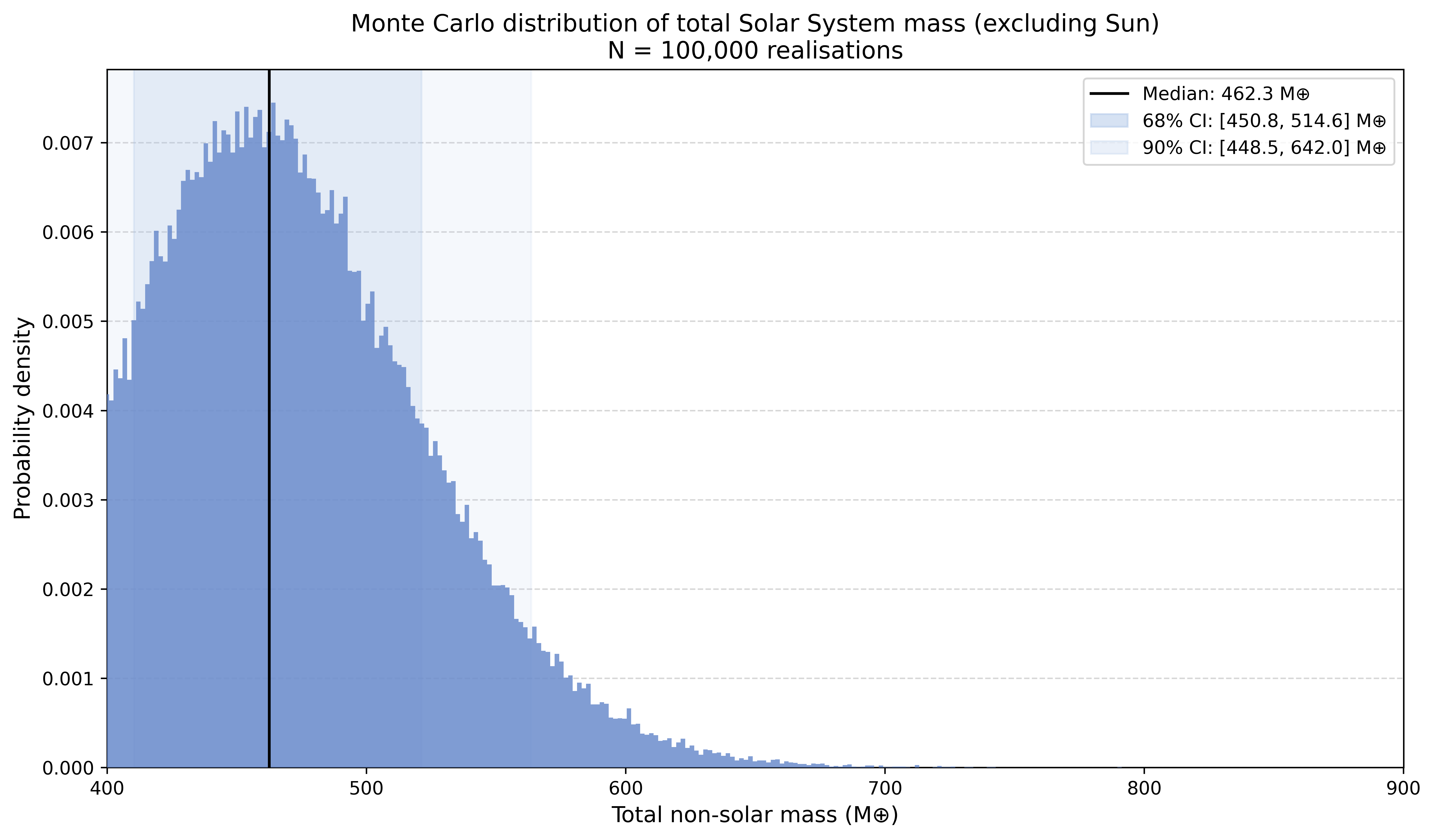}
\caption{Monte Carlo probability distribution of the total non-solar
mass of the Solar System ($N = 10^5$ realisations). It is shown as a smoothed kernel density estimate of the Monte Carlo samples. The distribution is
strongly right-skewed, with a median of $462.4\,\Mearth$, 68\% credible
interval $[451,\,515]\,\Mearth$ (dark shading), and 90\% credible
interval $[449,\,642]\,\Mearth$ (light shading). The asymmetry is a
direct consequence of the log-normal uncertainty in the Hills cloud mass.}
\label{fig:distribution}
\end{figure}

\begin{figure}[h]
\centering
\includegraphics[width=0.90\textwidth]{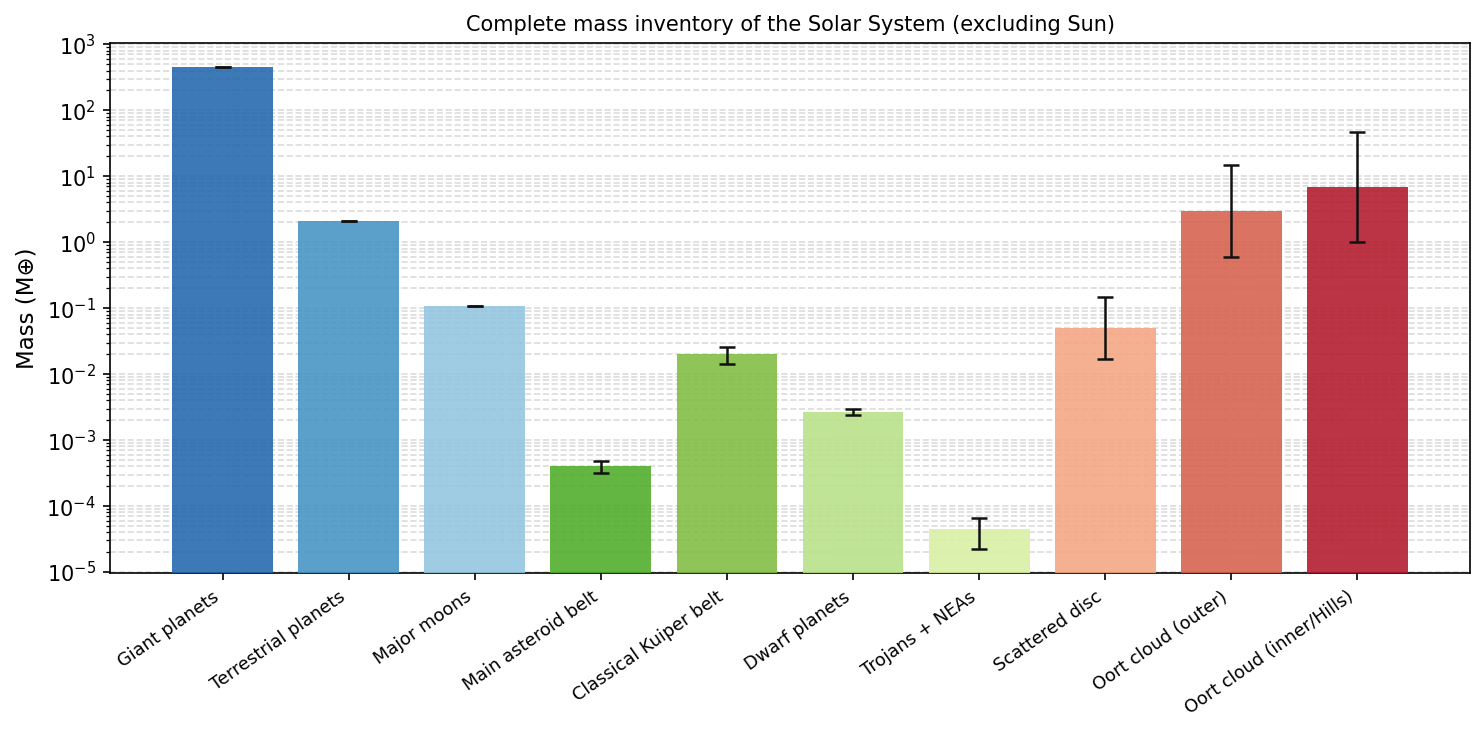}
\caption{Complete mass inventory of the Solar System on a single
logarithmic bar chart, spanning eleven orders of magnitude from Jupiter
($444.6\,\Mearth$) to the small satellite and ring populations
($\sim$$5 \times 10^{-6}\,\Mearth$). Colour progression from dark blue
(giant planets) to light green (inner trans-Neptunian) to orange/red
(Oort cloud) reflects increasing uncertainty.}
\label{fig:fullinventory}
\end{figure}

The full probability distribution (Figure~\ref{fig:distribution}) is
strongly right-skewed, with a sharp lower cutoff near $\sim$$448\,\Mearth$
and a long tail extending beyond $1000\,\Mearth$. The median is
$462.4\,\Mearth$, while the mean is substantially higher
($\sim$$490\,\Mearth$) because the distribution is not symmetric. The
median is the more robust summary statistic for a skewed distribution of
this kind and is adopted as our central value throughout.

\subsection{Planet Nine sensitivity}

If Planet Nine exists with a mass of $\sim$5--10\,$\Mearth$, it would
increase the central estimate by approximately 1--2\%, a shift well
within the existing Oort cloud uncertainty. Its existence or
non-existence does not materially affect our conclusions.

\subsection{Sensitivity analysis}

Replacing the inner Oort cloud log-normal factor from $f = 7$ to $f =
5$ narrows the 90\% CI to $[448,\,530]\,\Mearth$ but does not change the
median significantly. In all cases the qualitative conclusions are
unchanged: the planetary component is robustly $\sim$$447\,\Mearth$, and
the uncertainty is entirely attributable to the Oort cloud.

\section{Discussion}
\label{sec:discussion}

\subsection{The current inventory as a test of Solar System formation models}

The mass inventory compiled in Section~\ref{sec:results} encodes, in
the deficit between current and primordial masses, a quantitative record
of the dynamical violence of the early Solar System.

The Nice Model \citep{Tsiganis2005, Gomes2005, Morbidelli2005} requires
a primordial trans-Neptunian disc of approximately $35\,\Mearth$ between
$\sim$15 and 35\,AU, subsequently disrupted with $\sim$99\% of its mass
scattered or ejected by dynamical instability. Our inventory finds a
current trans-Neptunian mass (excluding the Oort cloud) of
$\sim$$0.07\,\Mearth$. The ratio of primordial to current trans-Neptunian
mass is therefore approximately 500:1, consistent with the Nice Model
prediction.

A critical subtlety is that not all ejected primordial disc mass was
lost from the Solar System. A fraction was deposited into the Oort cloud
during the dynamical instability: models suggest that roughly 1--5\% of
the disc mass was captured into the cloud, implying an Oort cloud mass
of 0.3--1.8\,$\Mearth$ from this channel alone, consistent with the
lower end of our adopted range. The large upper end of our Oort cloud
estimate ($\sim$$100\,\Mearth$) would require additional mass capture
from the Sun's birth cluster and is in mild tension with Nice Model
formation efficiencies.

The Grand Tack hypothesis \citep{Walsh2011} predicts an original
asteroid belt mass of $\sim$$1\,\Mearth$, reduced to the present
$\sim$$4 \times 10^{-4}\,\Mearth$ --- a depletion factor of roughly
2500. Our measurement is entirely consistent with this prediction.

\subsection{Where did the mass go? A budget of losses}

Table~\ref{tab:budget} constructs an approximate budget of the
$\sim$$35\,\Mearth$ of primordial trans-Neptunian disc material.

\begin{table}[h]
\centering
\caption{Approximate mass budget of the primordial trans-Neptunian disc.}
\label{tab:budget}
\begin{tabular}{lcc}
\toprule
\textbf{Destination} & \textbf{Current mass ($\Mearth$)} & \textbf{Fraction of original $\sim$$35\,\Mearth$} \\
\midrule
Classical Kuiper belt          & 0.020              & $\sim$0.06\%  \\
Scattered disc                 & $\sim$0.05         & $\sim$0.14\%  \\
Oort cloud (central estimate)  & $\sim$10           & $\sim$29\%    \\
Ejected to interstellar space  & $\sim$25 (by diff.)& $\sim$71\%    \\
\bottomrule
\end{tabular}
\end{table}

This budget makes explicit a point not always stated clearly in the
literature: the Oort cloud, if our central estimate is correct, retains
more mass from the primordial trans-Neptunian disc than all currently
observable small-body populations combined, by a factor of $\sim$100.
The vast majority of disc mass was ejected entirely from the Solar System
and is now permanently inaccessible to observation.

\subsection{Comparison with exoplanetary systems and protoplanetary discs}

ALMA surveys of protoplanetary discs around Sun-like stars in nearby
star-forming regions find median disc dust masses of only 20--30\% of
the total mass of exoplanet systems around similar stars, raising the
question of whether discs are systematically underestimated or whether
planet formation efficiency is extremely high. 

Our Solar System fits
naturally into this picture: the total mass currently retained in
planets and small bodies (excluding the Oort cloud) is approximately
$447.8\,\Mearth$. This value represents the total gravitational mass (gas + ices + refractory material), not solely the condensed solid component, with small bodies contributing only $\sim$$0.11\,\Mearth$. This is consistent with a high planet formation efficiency.

A direct comparison with ALMA disc masses is complicated by the fact
that ALMA measures dust mass at millimetre wavelengths, requiring
assumptions about the dust-to-gas ratio and grain size distribution. As
a rough check, the mean dust mass in the Lupus star-forming region
(age $\sim$1--3\,Myr) is approximately 0.2--0.4\,$\Mearth$ for the dust
component alone \citep{Zhang2015} --- broadly consistent with the
minimum-mass solar nebula estimate and with the primordial disc mass
implied by Nice Model simulations.

\subsection{The Oort cloud uncertainty as a scientific priority}

The central quantitative result of this paper --- that 98\% of the
variance in the total Solar System mass comes from the inner Oort cloud
alone --- has a direct practical implication: any significant improvement
in our knowledge of the Solar System's total non-solar mass requires,
above all, a better constraint on the Oort cloud.

The Vera C.\ Rubin Observatory (LSST) will survey the sky to
unprecedented depth over the next decade, expanding the known population
of trans-Neptunian objects at distances up to $\sim$100\,AU and improving
our census of Hills cloud (sednoid-class) objects. This will potentially
reduce the Hills cloud uncertainty factor from $\sim$7 to $\sim$3,
narrowing the 90\% CI on the total mass from $[448,\,642]\,\Mearth$ to
approximately $[449,\,480]\,\Mearth$. Improved modelling of the
long-period comet flux with ZTF and JWST could reduce the outer Oort
cloud uncertainty from our current factor of 5 to perhaps a factor of
2--3 within a decade.

Even in the best foreseeable scenario, the Oort cloud will remain the
dominant uncertainty in the Solar System mass budget for the next
several decades. The honest acknowledgment of this irreducible
uncertainty is, we argue, itself a scientifically useful contribution.

\section{Conclusions}
\label{sec:conclusions}

We have compiled the first systematic mass inventory of the Solar System
excluding the Sun, combining spacecraft-derived measurements, planetary
ephemeris analyses, and population surveys into a single self-consistent
table with propagated uncertainties. Our main results are the following.

The total non-solar mass of the Solar System is $462\,\Mearth$ (median),
with a 68\% credible interval of $[451,\,515]\,\Mearth$ and a 90\%
credible interval of $[449,\,642]\,\Mearth$. The extreme asymmetry of
this interval is a direct consequence of the log-normal uncertainty in
the Oort cloud mass.

The mass budget is overwhelmingly dominated by the giant planets, which
contribute $444.6\,\Mearth$, or 96.2\% of the median total. Jupiter
alone accounts for 68.7\% of all non-solar mass. All small-body
populations combined --- asteroid belt, Kuiper belt, scattered disc,
sednoids, and dwarf planets --- contribute approximately $0.08\,\Mearth$
excluding the Oort cloud.

A Monte Carlo variance decomposition demonstrates that 98\% of the total
uncertainty in the Solar System mass budget is attributable to a single
component: the inner Oort cloud (Hills cloud), for which no direct
observational constraints currently exist.

Comparison with Solar System formation models shows that the current
small-body inventory retains only $\sim$0.2\% of the primordial
trans-Neptunian disc mass postulated by the Nice Model, and our central
Oort cloud estimate of $\sim$$10\,\Mearth$ is consistent with a
formation efficiency of $\sim$29\% from that primordial disc.

Any future improvement in the total mass budget of the Solar System will
require, above all, better constraints on the Oort cloud mass. The Vera
C.\ Rubin Observatory and continued long-period comet surveys offer the
most promising near-term paths toward reducing the Hills cloud
uncertainty.

The inventory table, Monte Carlo code, and figures presented in this
work are intended as a baseline reference for future studies of Solar
System mass budgets, formation model comparisons, and exoplanetary
system analogues.

\bibliographystyle{aasjournal}

\end{document}